\documentclass[%
reprint,
superscriptaddress,
showpacs,preprintnumbers,
nofootinbib,
amsmath,amssymb,
aps,
prl,
]{revtex4-2}

\usepackage{mathtools, braket}
\usepackage{graphicx}  
\usepackage{bm}  
\usepackage{hyperref}  
\graphicspath{{figures/}}   

\usepackage{booktabs}
\usepackage{multirow} 
\newcommand{\cellbreak}[2][c]{\begin{tabular}[#1]{@{}c@{}}#2\end{tabular}}
\usepackage{changes}
\usepackage{todonotes}
\usepackage{subcaption}

\usepackage{siunitx} 
\DeclareSIUnit{\ohm}{\Omega}
\newcommand{\AlOx}{$ \text{AlO}_x $\@ }

\AtBeginDocument{
	\heavyrulewidth=.08em
	\lightrulewidth=.05em
	\cmidrulewidth=.03em
	\belowrulesep=.65ex
	\belowbottomsep=0pt
	\aboverulesep=.4ex
	\abovetopsep=0pt
	\cmidrulesep=\doublerulesep
	\cmidrulekern=.5em
	\defaultaddspace=.5em
}

\binoppenalty=\maxdimen
\relpenalty=\maxdimen

\makeatletter
\renewcommand\@makecaption[2]{%
	\par
	\vskip\abovecaptionskip
	\begingroup
	\rmfamily
	\begingroup
	\samepage
	\flushing
	\let\footnote\@footnotemark@gobble
	\@make@capt@title{#1}{#2}\par
	\endgroup
	\endgroup
	\vskip\belowcaptionskip
}
\makeatother

\begin{document}
	\title{Fluxons in high-impedance long Josephson junctions}
	\author{Micha Wildermuth}
	\author{Lukas Powalla}
	\author{Jan Nicolas Voss}
	\author{Yannick Sch\"on}
	\author{Andre Schneider}
	\affiliation{Institute of Physics, Karlsruhe Institute of Technology, 76131 Karlsruhe, Germany}
	\author{Mikhail V. Fistul}
	\affiliation{National University of Science and Technology MISIS, Moscow 119049, Russia}
	\affiliation{Theoretische Physik III, Ruhr-Universit\"at Bochum, 44801 Bochum Germany} 
	\author{Hannes Rotzinger}
	\email{Rotzinger@kit.edu}
	\affiliation{Institute of Physics, Karlsruhe Institute of Technology, 76131 Karlsruhe, Germany}
	\affiliation{Institute of Quantum Materials and Technology, Karlsruhe Institute of Technology, 76131 Karlsruhe, Germany}
	\author{Alexey V. Ustinov}	
	\affiliation{Institute of Physics, Karlsruhe Institute of Technology, 76131 Karlsruhe, Germany}
	\affiliation{National University of Science and Technology MISIS, Moscow 119049, Russia}	
	\affiliation{Institute of Quantum Materials and Technology, Karlsruhe Institute of Technology, 76131 Karlsruhe, Germany}
	\affiliation{Russian Quantum Center, Skolkovo, Moscow 143025, Russia}
	
	\date{\today}
	
	
	\begin{abstract}
		The dynamics of fluxons in long Josephson junctions is a well-known example of soliton physics and allows for studying highly nonlinear relativistic electrodynamics on a microscopic scale.
		Such fluxons are supercurrent vortices that can be accelerated by a bias current up to the Swihart velocity, which is the characteristic velocity of electromagnetic waves in the junction. 
		We experimentally demonstrate slowing down relativistic fluxons in Josephson junctions whose bulk superconducting electrodes are replaced by thin films of a high kinetic inductance superconductor. 
		Here, the amount of magnetic flux carried by each supercurrent vortex is significantly smaller than the magnetic flux quantum $ \Phi_0 $. 
		Our data show that the Swihart velocity is reduced by about one order of magnitude compared to conventional long Josephson junctions. 
		At the same time, the characteristic impedance is increased by an order of magnitude, which makes these junctions suitable for a variety of applications in superconducting electronics.
	\end{abstract}
	
	
	\maketitle
	
	
	The Josephson effect and weak links in superconductors \cite{Josephson1964, Josephson1965} are at the basis of a wide range of applications within superconducting electronics and many related fields. 
	The well-known examples are superconducting quantum interference devices \cite{Jaklevic1964QuantumInterferenceEffectsInJosephsonTunneling, Zimmerman1970DesignAndOperationOfStableRFbiasedSuperconductingPointContactQuantumDevices, Mercereau1970SuperconductingMagnetometers, Clarke1966SuperconductingGalvanometerEmployingJosephsonTunnelling}, voltage standard circuits \cite{Taylor1967, Field1973, Hamilton2000JosephsonVoltageStandards}, and superconducting qubits \cite{Makhlin1999JosephsonJunctionQubitsWithControlledCouplings, Clarke2008SuperconductingQuantumBits}.
	The dynamics of charges and electromagnetic fields in Josephson junctions (JJs) is governed by the phase difference between the overlapping wave functions of superconducting electrodes \cite{Josephson1964, Josephson1965}. 
	In spatially extended JJs, the phase difference can vary in both time and space, which gives rise to a variety of propagating electromagnetic excitations. 
	Common examples are linear waves formed by plasma oscillations of the Cooper pair density (Josephson plasmons), particle-like nonlinear wave packets with conserved amplitude, shape, and velocity (solitons) \cite{Barone1971TheoryAndApplicationsOfTheSineGordonEquation, Scott1973Soliton, Scott1976MagneticFluxPropagationOnAJosephsonTransmissionLine, McLaughlin1978, Lomdahl1985SolitonsInJosephsonJunctions}, as well as their bound states formed by soliton-antisoliton pairs oscillating around their common center of mass (breathers) \cite{Kaup1978SolitonsAsParticlesOscillatorsAndInSlowlyChangingMedia, Kivshar1989DynamicsOfSolitonsInNearlyIntegrableSystems}.
	
	In Josephson junctions, solitons occur in the form of Josephson vortices, often called fluxons \cite{Scott1973Soliton, Scott1976MagneticFluxPropagationOnAJosephsonTransmissionLine, McLaughlin1978, Ustinov1998}, which are pinned at the tunnel barrier plane and may propagate along this plane \cite{Kulik1967WavePropagation, Scott1976MagneticFluxPropagationOnAJosephsonTransmissionLine, Matsuda1983FluxonPropagationOnAJosephsonTransmissionLine}.
	By applying a bias current across the junction, these vortices can be accelerated up to the speed of light inside the Josephson transmission line, which is noted as Swihart velocity $ \bar{c} $ \cite{Swihart1961}.
	The vortex's supercurrent is associated with a spatially localized $ 2 \pi $-kink in the superconducting phase difference across the junction.
	In ``conventional'' JJs, bulk electrodes provide complete magnetic screening, so that the fluxoid quantization of the phase in $ 2 \pi $ units is linked to the magnetic flux carried by the vortex, which in turn is quantized in units of the magnetic flux quantum $ \Phi_0 = h/2e $ \cite{Josephson1964, Josephson1965}.
	
	The system's properties such as velocity and spatial extension of a fluxon are governed by the tunnel barrier's capacitance $ C $ and critical current density $ j_\text{c} $ as well as the lead inductance $ L_0 $ along the propagation direction.
	The precise controllability of these parameters qualifies Josephson vortices as excellent candidates for quantitative exploration of soliton physics.
	A preferred toy model is a quasi one-dimensional long Josephson junction (LJJ), whose length $ \ell $ exceeds the characteristic spatial scale of the vortex $ \lambda_\text{J} $, whereas the width $ w $ is much smaller than $ \lambda_\text{J} $.
	Extensive experiments in the past demonstrated, for instance, soliton-(anti)soliton interactions \cite{Ustinov1998, Fujimaki1987Spatiotemporal}, interplay with cavity resonances \cite{Chen1971, Fulton1973, Fiske1964, Kulik1965TheoryOfStepsOfVoltageCurrentCharacteristicOfTheJosephsonTunnelCurrent, Kulik1967TheoryOfResonancePhenomenaInSuperconductingTunneling}, Lorentz contraction of relativistic solitons \cite{Matsuda1983FluxonPropagationOnAJosephsonTransmissionLine, Laub1995LorentzContractionOfFluxQuantaObservedInExperimentsWithAnnularJosephsonTunnelJunctions}, and flux-flow dynamics of dense chains of solitons \cite{Yoshida1978FluxFlowCharacteristicsOfALargeJosephsonJunction, Nakajima1978DynamicVortexMotionInLongJosephsonJunctions}. 
	The latter regime finds its applications in microwave and millimeter-wave generation \cite{Nagatsuma1983Flux, Koshelets2001RadiationLinewidthOfFluxFlowOscillators} and amplification of microwave signals \cite{Nagatsuma1985FluxFlowTypeJosephsonLinearAmplifier, Nordman1995SuperconductiveAmplifyingDevicesUsingFluxonDynamics}.
	
	In all previous experiments with conventional JJs, the typical Swihart velocity was about a few percent of the light velocity in vacuum, while the junction's characteristic impedance was typically less than a few Ohms \cite{Barone1982, Likharev1986, Holdengreber2019Impedance}.
	These parameters are limited by the electrode's geometric inductance, which is given by the magnetic field penetration depth in the bulk superconducting electrodes and confined by the feasible structure size. 
	In particular, the very low characteristic impedance of LJJs remained the major obstacle limiting their applications in superconducting electronics.
	
	In this work, we overcome the above constraints by at least one order of magnitude via replacing the bulk electrodes of LJJ with a high kinetic inductance superconductor (HKIS), which increases the total inductance of the Josephson transmission line beyond the purely geometrical limit.
	Using the sine-Gordon model, we evaluate the impact of kinetic inductance on the Swihart velocity, Josephson length, and junction impedance.
	We verify these predictions by transport measurements at different magnetic fields, temperatures, and under microwave irradiation. 
	We demonstrate a reduction of the Swihart velocity by one order of magnitude compared to the conventional junctions. 
	Correspondingly, we estimate the characteristic impedance of our junctions to be a few tens of Ohms, opening the way towards matching them to standard 50-Ohm microwave cables and circuits.
	\begin{figure}[t]
		\includegraphics[scale=1.0]{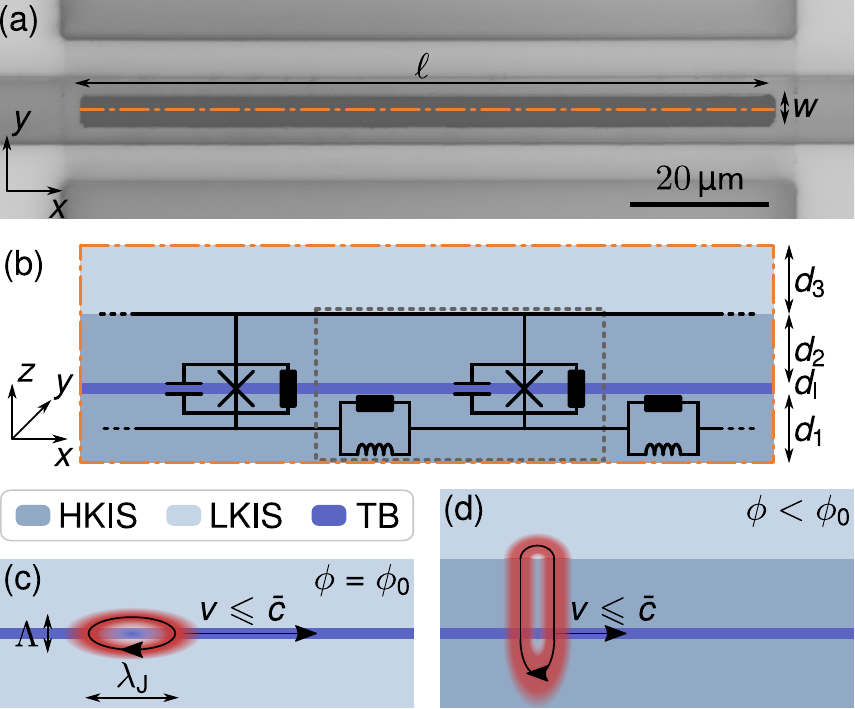}
		\caption{
			(a) Micrograph of a long junction (dark gray area) in quasi-overlap geometry in the top view and
			(b) schematic cross-section of the junction stack (along the dash-dotted line). 
			The junction consists of layers of a high kinetic inductance superconductor (HKIS), an insulating tunnel barrier (TB), and a HKIS proximitized by low kinetic inductance superconductor (LKIS). 
			The equivalent circuit of the LJJ's unit cell (dashed gray line) consists of a resistively and capacitively shunted junction across the TB together with resistive and inductive leads.
			(c) Josephson vortices (schematically shown as reddish ring current) arise in conventional Josephson junction with bulk LKISs as fluxons, each of them carrying one magnetic flux quantum $ \Phi_0 $.
			(d) In impedance-tailored junctions including HKISs, Josephson vortices appear as fluxoids, which have reduced length, speed, and magnetic flux.}
		\label{fig:JJ-scheme}
	\end{figure}

	Conventional LJJs can theoretically be modeled by lumped elements of resistively and capacitively shunted junctions \cite{Josephson1964, Stewart1968CurrentVoltageCharacteristicsOfJosephsonJunctions, McCumber1968EffectOfACImpedanceonDCVoltageCurrentCharacteristicsOfSuperconductorWeakLinkJunctions} in $ z $-direction, which are extended along the $ x $-axis and thus connected via inductive leads. 
	A finite resistance in parallel to an inductor in the equivalent scheme is due to surface losses \cite{Scott1964DistributedDeviceApplicationsOfTheSuperconductingTunnelJunction}, see Fig.\@ \ref{fig:JJ-scheme} (b).
	The resulting perturbed sine-Gordon equation \cite{Barone1971TheoryAndApplicationsOfTheSineGordonEquation, Ustinov1998}
	\begin{equation}\label{eq:sine-Gordon}
		\partial_{\tau\tau}\varphi - \partial_{\chi\chi}\varphi + \sin\varphi = \gamma - \alpha\partial_\tau\varphi + \beta\partial_{\chi\chi\tau}\varphi
	\end{equation}
	describes the junction's phase dynamics $ \varphi(\chi, \tau) $.
	The time $ t $ and the spatial coordinate $ x $ are normalized to $ \tau = \omega_\text{p}t $ and $ \chi = x/\lambda_\text{J} $, respectively, with the Josephson plasma frequency $ \omega_\text{p} = \left(2\pi j_\text{c}/c\Phi_0\right)^{1/2} $ as inverse time scale and the Josephson penetration length $ \lambda_\text{J} = \left(\Phi_0/2\pi j_\text{c}L_0^\square\right)^{1/2} $ as the characteristic length.
	Here $ j_\text{c} $ denotes the critical current density of the tunnel barrier, $ c = C/lw $ its specific capacitance, and $ L_0^\square $ the lead inductance per square.
	The left side of the perturbed sine-Gordon equation \eqref{eq:sine-Gordon} is a wave equation, which describes the Josephson transmission line with the characteristic (Swihart) velocity $ \bar{c} = \lambda_\text{J}\omega_\text{p} = (c L_0^\square)^{-1/2} $ \cite{Swihart1961}.
	The terms on the right side of Eq.\@ \eqref{eq:sine-Gordon} denote perturbations, namely a normalized bias current density $ \gamma = j_\text{b} /j_\text{c} $, ohmic dissipation due to quasiparticle tunneling $ \alpha $, and the surface resistance losses in the superconducting leads $ \beta $.
	
	The sine-Gordon model remains valid \cite{Alfimov1995} even with additional lumped elements of kinetic inductance $ L_\text{k} $ in the electrodes. 
	Here we compliment $ L_0^\square $ with a kinetic part.
	This additional kinetic inductance of the electrode material comes along with a larger magnetic field penetration depth $ \lambda_\text{L} $, which significantly modifies the vortex shape in such LJJs (see Fig. 1 (c) vs. Fig. 1 (d)). 
	The vortex current distributes inhomogeneously over the whole film thicknesses of the HKIS electrodes $ d_1 $, $ d_2 $.
	This  yields reduced effective participation of the bulk kinetic inductance to the Josephson length $ \lambda_\text{J} $.
	We take this effect into account by introducing a geometrical factor $ 0 < g(\vec{r}) < 1 $, such that for the junction's total lead inductance holds $ L_0^\square = L_\text{g}^\square + g(\vec{r})L_\text{k}^\square $.
	
	Compared to conventional long Josephson junctions, here the enlarged $ L_0^\square $ results in slower Swihart velocity $ \bar{c} \sim \left(L_0^\square\right)^{-1/2}$ and smaller vortex size $ \lambda_\text{J} \sim \left(L_0^\square\right)^{-1/2}$ \cite{Alfimov1995} while the junction impedance $ Z = \left(L_0^\square/c\right)^{1/2}/w $ \cite{Langenberg1966JosephsonTypeSuperconductingTunnelJunctionsAsGeneratorsOfMicrowaveAndSubmillimeterWaveRadiation}, correspondingly, increases.
	The lead inductance $ L_0^\square $ along $ z $ (see Fig.\@ \ref{fig:JJ-scheme} (b)) plays a minor role for supercurrent oscillations across the barrier, that is why the change in $ L_0^\square $ does not affect the Josephson plasma frequency, to the first order.
	Furthermore, a substantial fraction of the vortex's total $ 2\pi $ phase winding drops at the dominating kinetic inductance, which generates no magnetic field and results in incomplete magnetic screening.
	The phase winding (fluxoid) quantization remains valid, but it does no longer necessitate quantized magnetic flux. 
	The magnetic flux transported by a Josephson vortex $ \Phi$ is thus significantly smaller than $\Phi_0 $, so that this kind of vortex can be more correctly noted as ``fluxoid'' instead of ``fluxon''.
	Similar fluxoids were previously observed in arrays of JJs \cite{VanDerZant1991}, where the current distribution is predefined by the array geometry.
	Our approach to impedance-tailored LJJs provides fluxoids in a continuous Josephson medium where the current distribution evolves with no spatial constraints.
	\begin{figure}[t]
		\includegraphics[scale=1.0]{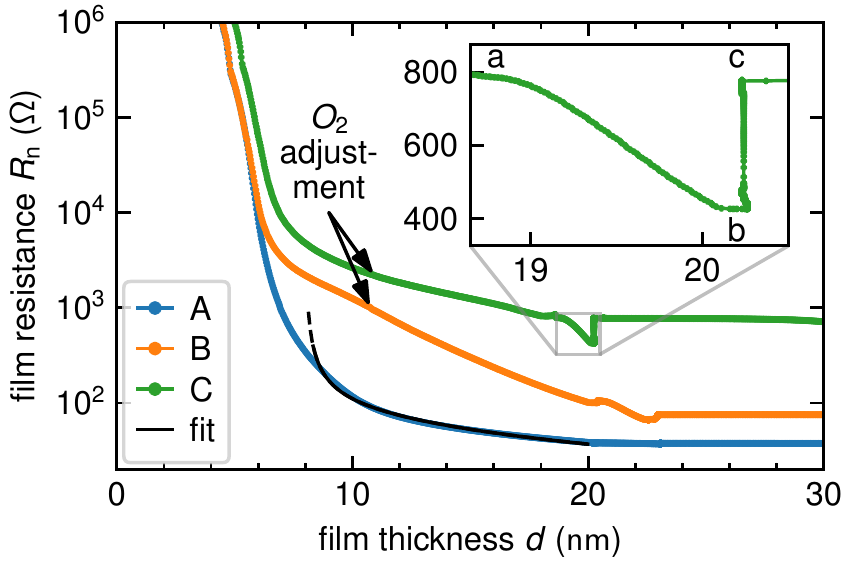}
		\caption{In-situ resistance monitoring during different trilayer depositions. 
			The fit of the thickness dependent normal resistance $ R_\text{n}(d) $ to the model of fine-grained polycrystalline thin films by Mayadas \textit{et al.}\@ \cite{Mayadas1969} allows us to estimate the final resistance and to readjust the oxygen partial pressure if needed (trilayers B and C). 
			The inset points out the tunnel barrier creation by deposition of pure Al (a to b) and subsequent static oxidation (b to c).}
		\label{fig:in-situ-resistance-measurement}
	\end{figure}
	
	The high kinetic inductance superconductor of our choice is a thin film of granular aluminum oxide ($ \text{AlO}_x $).
	Thin superconducting films of granular aluminum oxide have been used to establish macroscopic circuit quantum electrodynamics.
	This material consists of pure aluminum grains separated by intrinsic nanoscale tunnel barriers (TB) \cite{Deutscher1973}, which strongly influence both the normal and the superconducting transport properties.
	In the superconducting state, this granular material can be considered as a disordered network of Josephson junctions \cite{Deutscher1973}, each of them providing a kinetic-type Josephson inductance \cite{Josephson1964} related to the junction normal state tunnel resistance $ R_\text{n} $ and the superconducting gap $ \Delta $ by $ L_\text{k} = \hbar R_\text{n}/\pi\Delta $ \cite{Mattis1958, Glover1957, Annunziata2010TunableSuperconductingNanoinductors, Rotzinger2016, Maleeva2018CircuitQuantumElectrodynamicsOfGranularAluminumResonators}.
	The conductivity and inductance of HKIS formed by $\text{AlO}_x $ can vary over five orders of magnitude \cite{Deutscher1973GranularSuperconductingFilms, Chui1981}, depending on the oxygen concentration in the nanoscale TBs \cite{Ziemann1978}, which is controlled by the oxygen partial pressure during the reactive sputtering process \cite{Rotzinger2016}.
	This enormous versatility enables us using \AlOx for different purposes, e.g.,  for depositing junctions with an HKIS in the bottom electrode, for depositing an insulating TB, and for forming a top electrode as a combination of HKIS and pure aluminum, as illustrated in Fig.\@ \ref{fig:JJ-scheme} (b).
	As summarized in Tab.\@ \ref{tab:parameters}, we have fabricated three different junction stacks (A, B, and C) with varied values of $ L_\text{k}^\square $ and $ j_\text{c} $.
	
	Since the normal sheet resistance $ R_\text{n}^\square $ is the crucial parameter to obtain the desired kinetic inductance per unit square $ L_\text{k}^\square $, we monitor both the film thickness $ d $ and sheet resistance $ R_\text{n}^\square $ during the film deposition.
	This in-situ $ R(d) $ measurement allows us to fit the theoretical model for fine-grained polycrystalline thin films Mayadas \textit{et al.}\@ \cite{Mayadas1969} and to determine the specific resistance $ \rho_0$.
	For sample A, its value is $ \rho_0 = \SI{70.7 \pm 0.2}{\micro\ohm\centi\meter} $ yielding  $ R_\text{n}^\square \approx \SI{35}{\ohm} $ for a $ \SI{20}{\nano\meter} $ thick film (for details see Supplementary Material \ref{app:experimental-methods}).
	
	As the oxygen partial pressure can be adjusted during sputtering process, this kind of measurement is a powerful tool to achieve the aimed kinetic inductance value, with an accuracy of about $ \SI{10}{\percent} $, at a fixed film thickness.
	Figure\@ \ref{fig:in-situ-resistance-measurement} depicts such adjustments as knees and the creation of a tunnel barrier (emphasized in the inset).
	By reaching the targeted resistance at the end of the static oxidation process, we can assume the complete oxidation of the aluminum layer and also estimate the barrier thickness.
	
	The junctions were patterned from trilayers by using etching and anodic oxidization processes \cite{Gurvitch1983, Murduck1989NiobiumTrilayerProcessForSuperconductingCircuits}.
	The fabricated JJs were intentionally varied in length, width, and geometry. 
	The latter defines the distribution of the bias current over the junction \cite{Owen1967, Schwidtal1970} and therefore affects the vortex dynamics. 
	As discussed in detail in the Supplementary Material \ref{app:experimental-methods}, we fabricated junctions of square, inline and (quasi-)overlap geometries \cite{Sarnelli1991MagneticFieldDependenceOfTheCriticalCurrentInLongQuasiOverlapJosephsonJunctions}.
	We characterized the fabricated JJs (see Fig.\@ \ref{fig:JJ-scheme}(a)) by transport measurements at millikelvin temperatures and determined their characteristic parameters $ \lambda_\text{J} $, $ \bar{c} $ and $ \omega_\text{p} $ independently.
	\begin{table*}[tb]
		\caption{Properties of the fabricated trilayers and used geometry. 
			The normal conducting sheet resistance of the bottom electrode $ R_\square^\text{n} $ is extracted from the film deposition. 
			The critical current densities $ j_\text{c} $ are determined from squared junctions and Ambegaokar-Baratoff estimations coincide to switching current measurements with junction areas of $ (\SI{20}{\micro\meter})^2 $. 
			Magnetic thickness $ \Lambda^\text{exp} $ and Josephson penetration depth $ \lambda_\text{J} $ are derived from the magnetic field dependence of the critical current at $ T \approx \SI{300}{\milli\kelvin} $, from which also the geometry factor $ g(\vec{r}) $ follows. 
			The Swihart velocity $ \bar{c} $ and the impedance $ Z $ is acquired from the periodicity of zero-field and Fiske steps around $ \SI{1}{\kelvin} $.}
		\label{tab:parameters}
		\begin{tabular*}{\textwidth}{@{\extracolsep{\fill}}c S[table-format=3] S[table-format=2.2(1)] S[table-format=2(2)] S[table-format=2.1(2)] S[table-format=2(1)] S[table-format=1.2(2)] S[table-format=1.2(2)] S[table-format=2.2(1)] }
			\toprule
			{\cellbreak[t]{trilayer \\ class \\ geometry}} & {\cellbreak[t]{$ R_\square^\text{n} $ \\ $ (\si{\ohm}) $}} & {\cellbreak[t]{$ j_\text{c} $ \\ $ (\si{\ampere\per\centi\meter\squared}) $ \\ squared}} & {\cellbreak[t]{$ \Lambda^\text{exp} $ \\ $ (\si{\nano\meter}) $ \\ inline}} & {\cellbreak[t]{$ \lambda_\text{J} $ \\ $ (\si{\micro\meter}) $ \\ inline}} & {\cellbreak[t]{$ g(\vec{r}) $ \\ $ (10^{-2}) $ \\ inline}} & {\cellbreak[t]{$ \bar{c}_\text{zfs}/c_0 $ \\ $ (10^{-3}) $ \\ overlap}} & {\cellbreak[t]{$ \bar{c}_\text{FS}/c_0 $ \\ $ (10^{-3}) $ \\ overlap}} & {\cellbreak[t]{$ Z $ \\ $ (\si{\ohm}) $ \\ overlap}} \\
			\midrule
			A &  38 &  0.32 \pm 0.03 & 69 \pm 2 & \multicolumn{3}{l}{LJJ limit not reached for $ \ell \leq \SI{120}{\micro\meter} $} & 6.56 \pm 0.03 &  2.79 \pm 0.04 \\
			B &  75 & 12.5  \pm 0.3  & 72 \pm 9 & 17.2 \pm 2.2 & 15 \pm 4 & 4.27 \pm 0.06                                            & 3.64 \pm 0.02 &  4.11 \pm 0.06 \\
			C & 778 &  1.90 \pm 0.01 & 94 \pm 9 & 19.5 \pm 1.8 & 13 \pm 3 & 3.37 \pm 0.08                                            & 3.22 \pm 0.03 & 14.0  \pm 0.4  \\
			\bottomrule
		\end{tabular*}
	\end{table*}
	
	\begin{figure}[t]
		\includegraphics[scale=1]{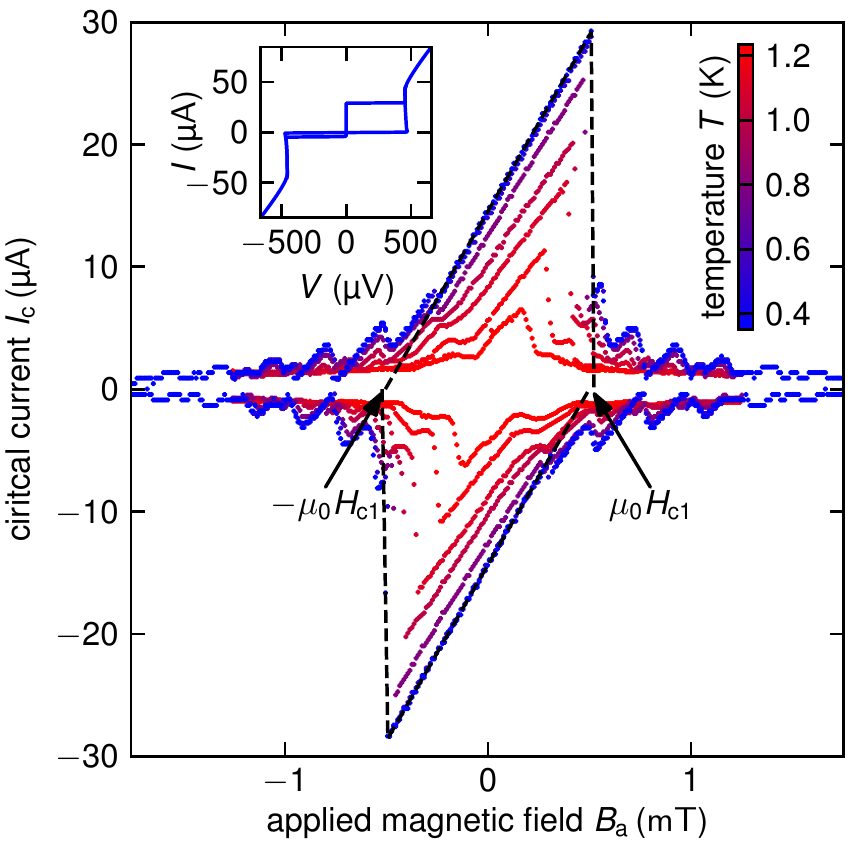}
		\caption{Magnetic diffraction pattern of a long junction of trilayer B in inline geometry at different temperatures.
			The linear decrease of the Meissner phase confirms the long junction limit, and the extrapolated root of the main lobe corresponds to $ \pm H_{\text{c}1} $.
			the asymmetric lobes arise because of inhomogeneously distributed bias currents and different electrode inductances \cite{Schwidtal1970, Barone1975CurrentflowInLargeJosephsonJunctions, Monaco2013}.
			The inset shows an IV-characteristic at the main maxima with large hysteresis implying high quality factors.}
		\label{fig:fraunhofer}
	\end{figure}
	
	In the first experiment, we determine the fluxoid's spatial extensions in both $ x $ and $ z $ direction, the Josephson length $ \lambda_\text{J} $ and the magnetic thickness $ \Lambda $ of the tunnel barrier from measurements of the dependence of the critical current on magnetic field applied in the plane of the tunnel barrier. 
	Examples of such critical current versus field patterns are depicted in Fig.\@ \ref{fig:fraunhofer}. 
	In high in-plane magnetic fields, where the junction is considered to be completely penetrated by magnetic flux along the $ x $ axis, $ \Lambda $ is determined from the critical current's periodicity $ \Delta B_\text{a} $ by $ \Lambda = \Phi_0/\ell\Delta B_\text{a} $.
	As can be seen in Tab.\@ \ref{tab:parameters}, $ L_\text{k} $ affects $ \Lambda $, since the proximitized top electrode's London penetration depth $ \lambda_\text{L} $ enlarges with increasing $ L_\text{k} $, whereas the bottom electrode of each stack is in the thin film limit $ d_1 \ll \lambda_{\text{L}1} $ and thus contributes to $ \Lambda $ with $ d_1/2 $ \cite{Weihnacht1969}.
	Together with the first critical field $ H_{\text{c}1} $, above which vortices can enter the junction, we calculate the vortex size $ \lambda_\text{J} = \Phi_0/\pi\mu_0 H_{\text{c}1}\Lambda $ \cite{Ferrell1963} and the kinetic inductance contributing locally to $ \lambda_\text{J} $.
	The comparison of this value $ g(\vec{r})L_\text{k}^\square $ with the kinetic inductance of the bottom layer $ L_\text{k}^\square $, estimated from the resistance $ R_\text{n}^\square $ measured in situ as $ R(d) $ during the sample deposition, yields the geometry factor $ g(\vec{r}) $ on the order of $ 10^{-1} $, as given in Tab.\@ \ref{tab:parameters}.
	
\begin{figure}
		\centering
		\includegraphics[scale=1]{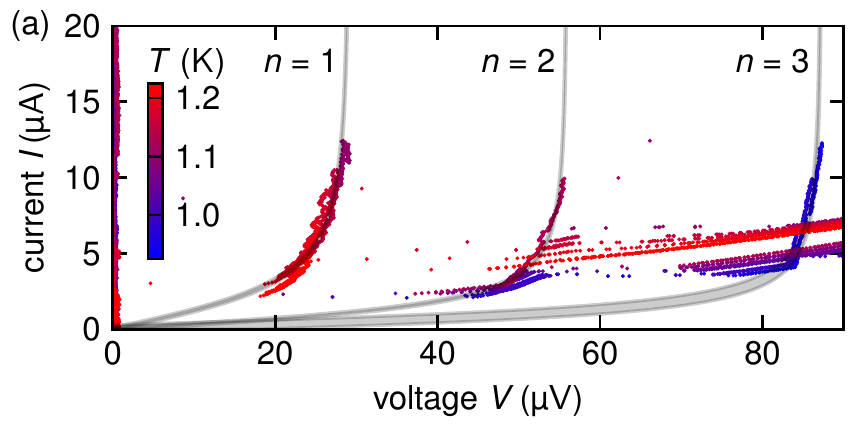}
		\vfill
		\includegraphics[scale=1]{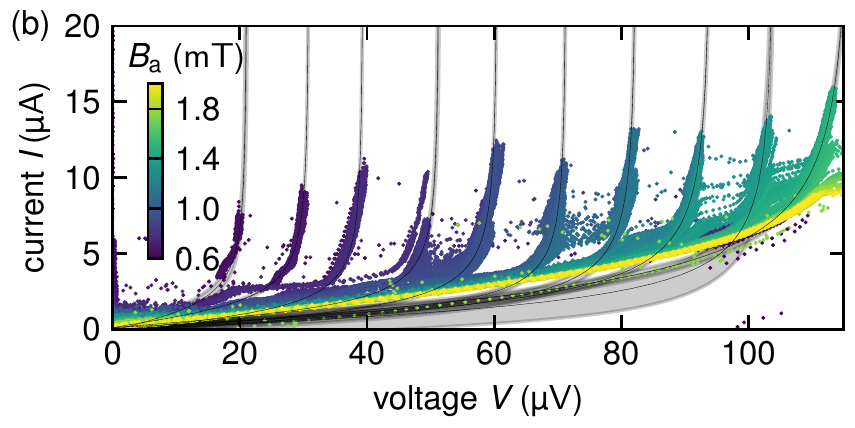}
		\caption{Current singularities of long junctions of sample B in quasi-overlap geometry without and with magnetic fields.
			The dark grey line and the grey shaded area display the fits to Eq.\@ \eqref{eq:current-steps} and their errors.
			(a) Zero-field steps arise only with sufficient damping in the junctions, which is realized by temperatures just below the critical temperature $ T_\text{c} \approx \SI{1.25}{\kelvin} $.
			(b) Fiske steps occur at different magnetic fields and their characteristic rounded shape originates from the increased damping at $ T = \SI{1.0}{\kelvin} $.}
		\label{fig:current-steps}
	\end{figure}

	In a second experiment, we determine the Swihart velocity from equidistant subgap current singularities originating in junction cavity mode excitations.
	In zero magnetic fields, these excitations are Josephson vortices, which are accelerated by the bias current, causing a Lorentz force, and reflected at the edges while reversing their polarity.
	Such resonant vortex oscillations manifest as current steps at integer multiples of the first zero-field step (ZFS) $ V_1^\text{ZFS} = \Phi_0\bar{c}/\ell $ \cite{Fulton1973}.
	Another kind of current singularities arises above the critical magnetic field where the Josephson frequency of a biased junction excites electromagnetic standing waves in the junction cavity.
	Such singularities are known as Fiske steps (FS) and occur at voltages with half the periodicity of ZFS $ V_1^\text{FS} = \Phi_0\bar{c}/2l $ \cite{Fiske1964, Kulik1965TheoryOfStepsOfVoltageCurrentCharacteristicOfTheJosephsonTunnelCurrent, Kulik1967TheoryOfResonancePhenomenaInSuperconductingTunneling}.
	As the vortex propagation velocity depends on the bias current $ \gamma $ and the damping parameter $ \alpha $, the characteristic shape of the $ n $th ZFS step is given by \cite{McLaughlin1978}
	\begin{equation}\label{eq:current-steps}
		V_n^{\text{ZFS}}(\gamma) = V_1^{\text{ZFS}}\frac{n}{\sqrt{1+\left(\frac{4\alpha}{\pi\gamma}\right)^2}}.
	\end{equation}
	The Swihart velocity $ \bar{c} $ is determined by the periodicity of the current singularities and the known junction length $ \ell $ (see Tab.\@ \ref{tab:parameters}). 
	As the junctions are underdamped (note the large hysteresis between critical and retrapping currents in the IV characteristics in the inset of Fig.\@ \ref{fig:fraunhofer}), for reliably observing these current 
	singularities arising from the subgap resistance branch it helps to increase the damping by increasing temperature of the sample. 
	Then, however, the Stewart-McCumber branch cuts the lower part of the higher-order steps, as shown in Fig.\@ \ref{fig:current-steps}. 
	For underdamped junctions, cavity oscillations are unstable for $ \omega \lesssim \omega_\text{p} $ \cite{Cirillo1997, Cirillo1998}, which explains missing the first FS in Fig.\@ \ref{fig:current-steps}.
	
	In the third experiment, we determine the Josephson plasma frequency $ \omega_\text{p} $ by measuring the plasma resonance of a square-shaped junction made of the trilayer with the tunnel barrier TB. 
	The Josephson plasma oscillations are excited by applying external microwave irradiation. 
	Resonant, subharmonic, or superharmonic driving \cite{Gronbech2004} causes a multi-valued switching current from the zero to the nonzero voltage state.
	The secondary peaks in the switching current distribution \cite{Wallraff2003, Gronbech2004, Blackburn2010} (see inset of Fig.\@ \ref{fig:plasma_frequency} in \ref{app:plasma-frequency}) are identified as resonant currents, for which the fixed external drive frequency equals the Josephson plasma resonance frequency $ \omega_{0} $, its integer multiples, or its fractions of $ \omega_{0} $.
	Since the bias current tilts the washboard potential of a JJ \cite{Martinis1985} and thus affects its shape, the associated internal oscillation frequency holds $ \omega_{0}(\gamma) = \omega_\text{p}\left(1-\gamma^2\right)^{1/4} $ \cite{Dahm1968}.
	Orthogonal distance regression, as shown in Fig.\@ \ref{fig:plasma_frequency}, yields the plasma frequency $ \omega_\text{p}/2\pi = \SI{13.28 \pm 0.05}{\giga\hertz} $, the critical current $ I_\text{c} = \SI{8.36 \pm 0.08}{\micro\ampere} $, and hence the specific tunnel barrier capacitance $ c = \SI{36.4 \pm 0.4}{\femto\farad\per\micro\meter\squared} $.
	
	To analyze the impact of the electrode's kinetic inductance on LJJs, their characteristic parameters, listed in Tab.\@ \ref{tab:parameters}, are compared with estimations for conventional LJJs with equal tunnel barrier properties $ j_\text{c} $ and $ c $, but made from pure aluminum. 
	Here we assume that pure aluminum electrodes have negligible kinetic inductance.
	The result of this comparison is that the electrode's kinetic inductance reduces both the Josephson length $ \lambda_\text{J} $ and the Swihart velocity $ \bar{c} $ by a factor of up to $ 40 $, while the Josephson plasma frequency $ \omega_\text{p} $ remains nearly unchanged. 
	Accordingly, the wave impedance of LJJs is increased by the same factor.
	The inductance contributing to Josephson plasma oscillations is dominated by the macroscopic stack TB rather than the nanoscopic TBs in \AlOx due to the much stronger intergrain coupling, so that the increase of $ L_\text{k} $ can be neglected to the first order.
	The combination of the independently measured parameters corresponds to the conventional sine-Gordon model with modified $ \bar{c} = \lambda_\text{J}\omega_\text{p} $.
	
	In conclusion, our results demonstrate a significantly reduced Swihart velocity in long Josephson junctions fabricated with high kinetic inductance electrodes. 
	In our work, we used disordered oxidized aluminum as a high kinetic inductance superconductor.
	Our experiments demonstrate a decrease in the vortex's size and a reduction of its limiting (Swihart) velocity by about one order of magnitude in comparison with conventional LJJs. 
	The measured Swihart velocities down to a small fraction of $ \num{3e-3} $ of the light velocity in vacuum, in turn, correspond to an increase junction's wave impedance up to $ \SI{14}{\ohm} $ compared to $ \SI{4} {\ohm} $  of conventional, similarly made LJJs.
	The high-kinetic inductance electrodes thus enable tailoring the junction impedance and facilitate solving the long-standing problem of impedance matching LJJs to external circuits and $ \SI{50}{\ohm} $ cables.
	Matching the impedance to external loads is crucial for increasing the efficiency of Josephson flux-flow oscillators used for microwave generation and amplification.
	Furthermore, the reduction of vortex size results in fewer charges participating in internal junction dynamics, a smaller effective capacitance over the vortex area $ C_\text{eff} $ and thus increases the effective charging energy $ E_\text{c,eff} = q^2/2C_\text{eff} $.
	As $ E_\text{c,eff} $ plays the key role in experimentally reaching the quantum regime of Josephson vortex dynamics \cite{Wallraff2003QuantumDynamicsOfASingleVortex}, high kinetic inductance electrodes also facilitate observing the quantum electrodynamics phenomena in long Josephson junctions.
	
	The authors are grateful for fruitful discussions with A. Shnirman, J. Lisenfeld, T. Wolz, and M. Spiecker. We also thank L. Radtke for his assistance during the sample fabrication. 
	The work was supported bilaterally by the German Science Foundation (DFG) through grant No. US 18/18-1 and the Russian Science Foundation through grant No. 19-42-04137. We also acknowledge support from the Landesgraduiertenf\"orderung of the state Baden-W\"urttemberg (M.W.), the Helmholtz International Research School for Teratronics (J.N.V.\@ and Y.S.\@), and the Carl Zeiss Foundation (A.S.\@), as well as partial support by the Ministry of Education and Science of the Russian Federation in the framework in the framework of the Program of Strategic Academic Leadership "Priority 2030" (M.V.F. and A.V.U.\@).

	\bibliographystyle{apsrev4-1}
	\bibliography{references}

	\clearpage
	\setcounter{secnumdepth}{2}
	\renewcommand{\thesection}{S\arabic{section}}
	\renewcommand{\thesubsection}{S\arabic{section}.\arabic{subsection}}
	\renewcommand{\thesubsubsection}{S\arabic{section}.\arabic{subsection}.\arabic{subsubsection}}
	\renewcommand{\thefigure}{S\arabic{figure}}
	\renewcommand{\theequation}{S\arabic{equation}}
	\setcounter{figure}{0}
	\setcounter{equation}{0}

	\onecolumngrid
	\begin{center}
		\textbf{\large Supplementary Material for ``Fluxons in high-impedance long Josephson junctions''}
	\end{center}
	\twocolumngrid

	\section{Experimental methods}\label{app:experimental-methods}
	The kinetic inductance of granular aluminum oxide ($ \text{AlO}_x $) can be estimated by $ L_\text{k}^\square = \hbar R_\text{n}^\square/\pi\Delta $ 
	\cite{Rotzinger2016}, where the superconducting gap $ \Delta $ is nearly constant for slightly different normal sheet resistances $ R_\text{n} $.
	The control of this $ R_\square^\text{n} $ is decisive to achieve the desired kinetic inductance \cite{Rotzinger2016}.
	For this reason, we monitor both the film thickness $ d $ and the normal resistance $ R_\text{n} $ during the pulsed DC magnetron sputter deposition of \AlOx.
	This measurement enables us to fit the specific conductance $ \sigma(d) \equiv \rho^{-1} = \left(R_\text{n}^\square d\right)^{-1} $ to estimate the final specific resistance $ \rho_{\text{n}0} $ in situ.
	To describe the thickness dependent specific conductance we use a model for fine-grained polycrystalline thin films by Mayadas \textit{et al.}\@ \cite{Mayadas1969}
	\begin{equation}\label{eq:Mayadas}
		\frac{\sigma}{\sigma_0} \equiv \frac{\rho_{\text{n}0}}{\rho_\text{n}} = 3\left[\frac{1}{3}-\frac{\alpha}{2}+\alpha^2-\alpha^3\ln\left(1+\frac{1}{\alpha}\right)\right].
	\end{equation}
	that solves a linearized Boltzmann equation concerning ordinary scattering mechanisms as in bulk materials and superimposed scattering at grain boundaries.
	Here, $ \sigma_0 $ denotes the intrinsic thickness independent conductivity from the film interior and $ \alpha \coloneqq \frac{l_0}{d}\frac{r}{1-r} $ is the ratio between the background mean free path $ l_0 $ and the film thickness $ d $, reduced by a scattering reflection coefficient $ r $.
	This estimation allows us to set the specific resistance accurately by readjusting the oxygen partial pressure if necessary and thus to achieve the desired sheet resistance $ R_\square^\text{n} $ at the fixed film thickness with a precision less than $ \SI{10}{\percent} $.
	The least-squares fit for trilayer A as shown in Fig.\@ \ref{fig:in-situ-resistance-measurement} allows an offset thickness, above which the model holds, and yields $ \rho_0 = \SI{70.7 \pm 0.2}{\micro\ohm\centi\meter} $ and $ \frac{l_0r}{1-r} = \SI{5.04 \pm 0.17}{\angstrom} $. 
	Assuming $ \rho_0 l_0 = 12\pi^3\hbar/e^2S_\text{F} = \SI{1.6e-11}{\ohm\per\centi\meter\squared} $ for \AlOx \cite{Cohen1968SuperconductivityInGranularAluminumFilms}, results in a $ l_0 = \SI{22.62 \pm 0.07}{\angstrom} $ and $ r = \SI{18.2 \pm 0.5}{\percent} $ that confirms diffusive transport in the granular material.
	The discrepancy of the measurement and the theoretical model for small film thicknesses originated in a inhomogeneous film thickness and the conductivity of the argon plasma that contributes especially for small film thicknesses, where the film is not entirely connected.
	
	The SIS tunnel junction stacks are grown on a c-plane sapphire substrate and patterned using photolithography and chlorine-based inductively coupled plasma etching.
	The junctions themselves are defined via anodic oxidation through a solvent of ammonium pentaborate in ethylene glycol and water, to isolate the top from the bottom electrodes.
	The leads to the upper aluminum electrodes are evaporated thermally, where the galvanic contact is ensured by previous argon milling \cite{Braumueller2015}.
	The junction designs vary in length ($ \SI{20}{\micro\meter} $ to $ \SI{120}{\micro\meter} $), width ($ \SI{2}{\micro\meter} $ to $ \SI{5}{\micro\meter} $) and geometry.
	
\begin{figure}[t]
		\centering
		\includegraphics[scale=1]{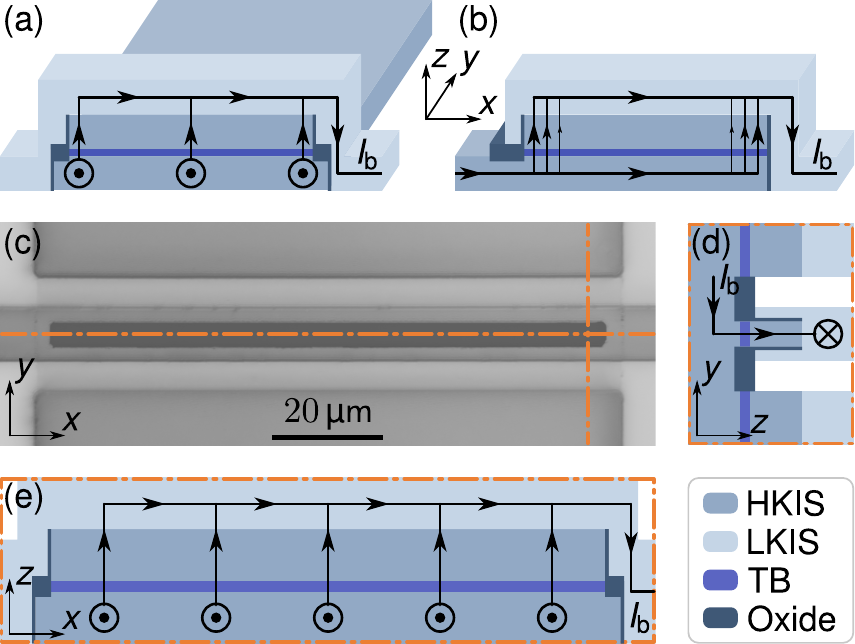}
		\caption{Bias current distribution in different junction geometries.
			(a) In the quasi-overlap geometry, the bias current distributes homogeneously over the bottom electrode made from a high kinetic inductance superconductor (HKIS) tunnels across the tunnel barrier (TB) and goes out via the low kinetic inductance superconductor (LKIS), which is isolated from the bottom electrode by an anodic oxidized layer (oxide).
			(b) In contrast, the inline geometry provides an inhomogeneous bias current distribution.
			(c) Micrograph of a long junction (dark gray area) in quasi-overlap geometry in the top view and schematic cross-sections along the (d) $ y $ and (e) $ x $ plane.
		}
		\label{fig:JJ-geometries}
	\end{figure}

	The junction geometry governs the distribution of bias currents over the junction, which drives the vortex (see Fig.\@ \ref{fig:JJ-geometries}).
	Since short squared junctions provide the most homogeneous current distribution, they are used to determine stack characteristic quantities such as critical current density and plasma frequency.
	The inline geometry provides an inhomogeneous bias current distribution and is suitable for magnetic diffraction patterns.
	To improve the homogeneity of the bias current, as desired for the investigation of soliton dynamics, a quasi-overlap geometry \cite{Sarnelli1991MagneticFieldDependenceOfTheCriticalCurrentInLongQuasiOverlapJosephsonJunctions} is used, where high kinetic inductance material is placed in the bias leads in junction vicinity.
	
	The low-temperature measurements are performed either in a $ ^3\text{He} $ cryostat or in a $ ^3\text{He}/^4\text{He} $ dilution refrigerator, whereby the DC measurement lines are low-pass filtered using combinations of CLC, RCR and copper powder filters \cite{Lukashenko2008ImprovedPowderFiltersForQubitMeasurements} at different temperature stages.
	Magnetic fields are applied in-plane and external fields are suppressed by a surrounding Cryoperm\textsuperscript{\textregistered} shield.

	\section{Low temperature measurements}\label{app:low-temperature-measurements}
	\subsection{Junction characterization}\label{app:junction-characterization}
\begin{figure}
		\centering
		\includegraphics[scale=1]{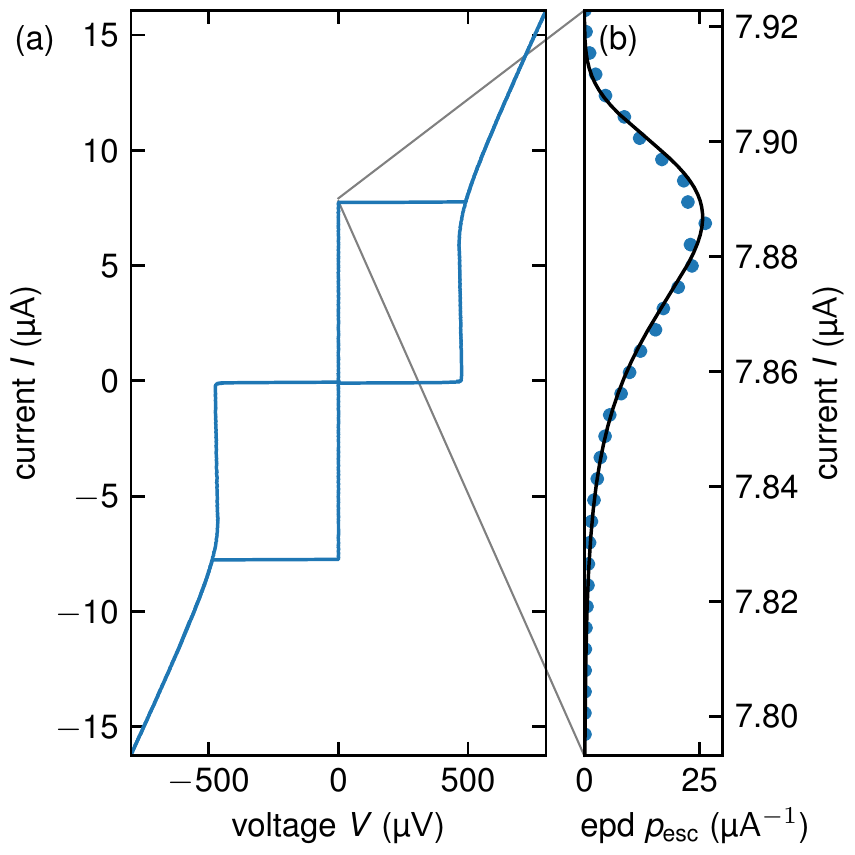}
		\caption{IV-characteristic and switching current distribution of a squared junction with $ (\SI{10}{\micro\meter})^2 $ of trilayer B at $ \SI{20}{\milli\kelvin} $.
			(a) The IV-characteristic shows a large hysteresis between the switching current from the zero-voltage to the nonzero-voltage state and the retrapping current due to low internal damping. 
			(b) The distribution of these switching currents yields the escape probability density (epd), which is fitted to Eq.\@ \eqref{eq:esc-pobability-density} including Eq.\@ \eqref{eq:thermal-escape-rate}.}
		\label{fig:switching_current}
	\end{figure}

	Transport measurements at low temperatures of about $ \SI{20}{\milli\kelvin} $ show IV-characteristics as in Fig.\@ \ref{fig:switching_current} (a). 
	The large hysteresis between switching and retrapping current confirms a high quality factor and only little quasiparticle excitations. 
	The critical currents $ I_\text{c} $ and the ciritcal current densities $ j_\text{c} = I_\text{c}/lw $ are determined by switching current measurements and the Ambegaokar-Baratoff relation. 
	In switching current measurements we detect the escape current from the zero-voltage to the nonzero-voltage state for $ 10\,000 $ events. 
	The integral equation for the escape probability \cite{Fulton1974LifetimeOfTheZeroVoltageStateInJosephsonTunnelJunctions}
	\begin{subequations}
		\begin{equation}\label{eq:esc-pobability}
			p(I)\mathrm{d}I = \Gamma(I)\left|\frac{\mathrm{d}I}{\mathrm{d}t}\right|^{-1} \left(1-\int\limits_0^I p(I^\prime)\mathrm{d}I^\prime\right)\mathrm{d}I
		\end{equation}
		can be solved for the escape probability density (epd), which reads
		\begin{equation}\label{eq:esc-pobability-density}
			p(I) = \Gamma(I)\left|\frac{\mathrm{d}I}{\mathrm{d}t}\right|^{-1}\exp\left(-\left|\frac{\mathrm{d}I}{\mathrm{d}t}\right|^{-1}\int\limits_0^I\Gamma(I^\prime)\mathrm{d}I^\prime\right).
		\end{equation}
		The epd depends on the current sweep rate $ \mathrm{d}I/\mathrm{d}t $ and the thermal activation rate \cite{Devoret1985MeasurementsOfMacroscopicQuantumTunnelingOutOfTheZeroVoltageStateOfACurrentBiasedJosephsonJunction}
		\begin{equation}\label{eq:thermal-escape-rate}
			\Gamma_\text{th}(I) = \frac{\omega_\text{p}}{2\pi}\exp\left(-\frac{E_\text{J}\frac{4\sqrt{2}}{3}(1-I/I_\text{c})^{3/2}}{k_\text{B}T}\right).
		\end{equation}
	\end{subequations}
	Here, the junction's potential is assumed as tilted washboard \cite{Stewart1968CurrentVoltageCharacteristicsOfJosephsonJunctions, McCumber1968EffectOfACImpedanceonDCVoltageCurrentCharacteristicsOfSuperconductorWeakLinkJunctions} with the Josephson energy $ E_\text{J} = \frac{\Phi_0 I_\text{c}}{2\pi } $. 
	From the fit of the epd of Eq.\@ \eqref{eq:esc-pobability-density} with Eq.\@ \eqref{eq:thermal-escape-rate}, shown in Fig.\@ \ref{fig:switching_current} (b), we can extract the critical current $ I_\text{c} $.
	For large squared junctions with areas of $ (\SI{20}{\micro\meter})^2 $, this result coincides with the Ambegaokar-Baratoff model \cite{AmbegaokarBaratoff1963}
	\begin{equation}\label{eq:Ambegaokar-Baratoff}
		I_\text{c} = \frac{\Delta_1(T)}{eR_\text{n}}K\left(\sqrt{1-\left(\frac{\Delta_1(T)}{\Delta_2(T)}\right)^2}\right),
	\end{equation}
	which estimates $ I_\text{c} $ from the normal resistance $ R_\text{n} $ and the two superconducting gaps $ \Delta_{1,2} $. 
	As typical for JJs with different superconductors the IV-characteristics show two effective gaps \cite{Nicol1960, Barone1982} at $ \Delta_{1,2}^\text{eff} = \frac{\Delta_{\text{AlO}_x} \pm \Delta_\text{Al}}{2} $ with the gap energies of proximitized aluminum $ \Delta_\text{Al} \simeq \SI{190}{\micro\electronvolt}\,..\,\SI{230}{\micro\electronvolt} $ and inversely proximitized disordered oxidized aluminum $ \Delta_{\text{AlO}_x} \simeq \SI{270}{\micro\electronvolt}\,..\,\SI{280}{\micro\electronvolt} $.

	\subsection{Plasma frequency}\label{app:plasma-frequency}
\begin{figure}
		\includegraphics[scale=1]{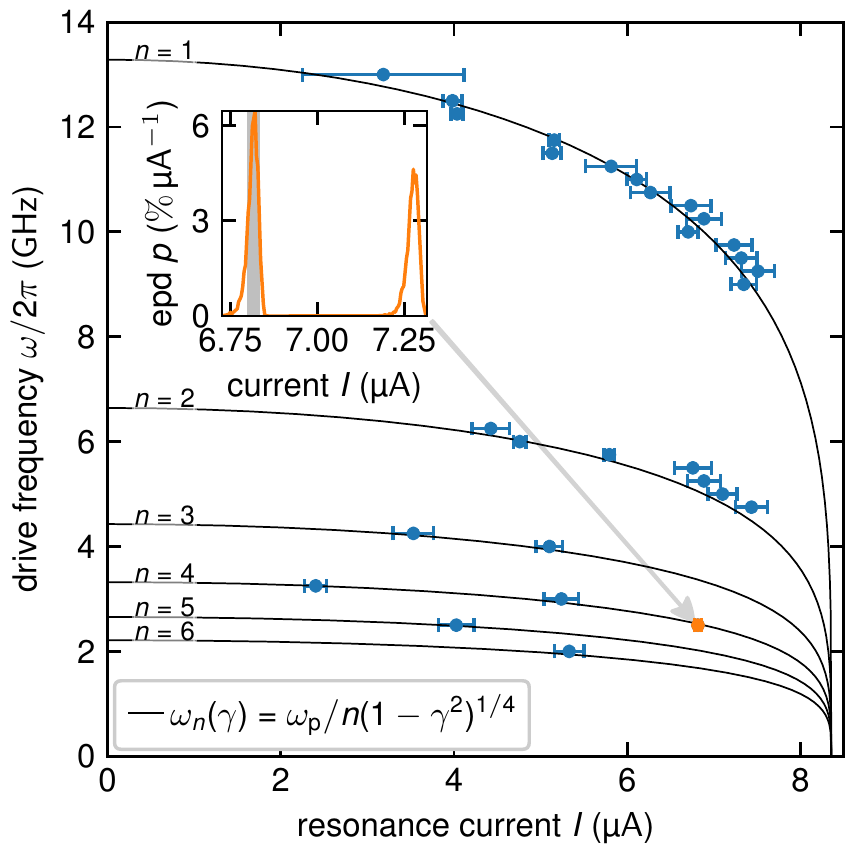}
		\caption{Resonance current at different external microwave drives of a squared junction with $ (\SI{10}{\micro\meter})^2 $ of trilayer B at $ \SI{20}{\milli\kelvin} $.
			If the internal resonance frequency or its $ n^\text{th} $ subharmonics matches the drive frequency, the switching current becomes multi-valued.
			The secondary peak in the switching current distribution is identified as resonance current and its full width half maximum as error.}
		\label{fig:plasma_frequency}
	\end{figure}

	In the washboard potential of a Josephson junction, thermally or microwave driven oscillations can excite the particle to the nonzero voltage state below the critical dc current $ I_\text{c} $ \cite{Devoret1985MeasurementsOfMacroscopicQuantumTunnelingOutOfTheZeroVoltageStateOfACurrentBiasedJosephsonJunction}. 
	Off-resonant microwave irradiation superimposes a small AC contribution to the DC bias and thus lowers the switching current continuously with increasing amplitude. 
	Resonant, sub- or superharmonic microwave irradiation, however, excites plasma oscillations that manifest as multi-valued switching current \cite{Wallraff2003, Gronbech2004, Blackburn2010}.
	As the internal resonance frequency $ \omega_{0} $
	\begin{equation}\label{eq:resonant-frequency}
		\omega_{0}(I) = \omega_\text{p}\left(1-\left(\frac{I}{I_\text{c}}\right)^2\right)^{1/4}
	\end{equation}
	depends on DC bias current $ I $, such a secondary peak can be assigned as a resonance current to a fixed drive frequency, as shown in Fig.\@ \ref{fig:plasma_frequency}.
	To determine the plasma frequency $ \omega_\text{p} = \left(2\pi I_\text{c}/\Phi_0 C\right)^{1/2} $, we take switching current distributions with $ 10\,000 $ events and irradiate various but fixed external drive frequencies with suitable drive power, so that two distinguishable peaks arise.
	Orthogonal distance regression yields a plasma frequency of $ \omega_\text{p}/2\pi = \SI{13.28}{\giga\hertz} $, a critical current of $ I_\text{c} = \SI{8.36}{\micro\ampere} $ and hence a specific capacitance of $ C/A = \SI{36.5}{\femto\farad\per\micro\meter\squared} $. 
	This is a typical number for tunnel barriers made from thermally oxidized aluminum \cite{Maezawa1995} and indicates that the additional kinetic inductance does not affect the plasma frequency in the first order.
	
\end{document}